\documentclass[10pt]{article}

\usepackage{amsmath}
\usepackage{amssymb}

\usepackage{graphicx}

\usepackage{cite}

\usepackage{color} 

\usepackage[labelfont=bf,labelsep=period,justification=raggedright]{caption}

\makeatletter
\renewcommand{\@biblabel}[1]{\quad#1.}
\makeatother

\date{}

\newcommand{\B}{\begin{eqnarray*}}
\newcommand{\E}{\end{eqnarray*}}
\newcommand{\Bnum}{\begin{eqnarray}}
\newcommand{\Enum}{\end{eqnarray}}
\newcommand{\bra}{\langle}
\newcommand{\ket}{\rangle}

\begin{document}

\begin{flushleft}
{\Large
\textbf{Mechanical Stress Inference for Two Dimensional Cell Arrays}
}
\\
\end{flushleft}
\begin{flushleft}
{\bf Kevin K. Chiou$^{1}$, 
Lars Hufnagel$^{2}$, 
Boris I. Shraiman$^{1,3,\ast}$}
\\
\ {1} Department of Physics, University of California, Santa Barbara, CA 91306, USA.
\\
\ {2} Department of Cell Biology and Biophysics, European Molecular Biology Laboratory, Heidelberg, Germany.
\\
\ {3} Kavli Institute for Theoretical Physics, University of California, Santa Barbara, CA 91306,USA.
\\
$\ast$ E-mail: shraiman@kitp.ucsb.edu
\end{flushleft}

\section*{Abstract}
Many morphogenetic processes involve mechanical rearrangement of epithelial tissues that is driven by precisely regulated cytoskeletal forces and  cell adhesion. The mechanical state of the cell and intercellular adhesion are not only the targets of regulation, but are themselves likely signals that coordinate developmental process. Yet, because it is difficult to directly measure  mechanical stress {\it in vivo} on sub-cellular scale, little is understood about the role of mechanics of development. Here we present an alternative approach which takes advantage of the recent progress in live imaging of morphogenetic processes and uses computational analysis of high resolution images of epithelial tissues to infer relative magnitude of forces acting within and between cells. We model intracellular stress in terms of bulk pressure and interfacial tension, allowing these parameters to vary from cell to cell and from interface to interface. Assuming that epithelial cell layers are close to mechanical equilibrium, we use the observed geometry of the two dimensional cell array to infer interfacial tensions and intracellular pressures. Here we present the mathematical formulation of the proposed Mechanical Inverse method and apply it to the analysis of epithelial cell layers observed at the onset of ventral furrow formation in the {\it Drosophila} embryo and in the process of hair-cell determination in the avian cochlea. The analysis reveals mechanical anisotropy in the former process and mechanical heterogeneity, correlated with cell differentiation, in the latter process. The method opens a way for quantitative and detailed experimental tests of models of cell and tissue mechanics.

\section*{Author Summary}
Mechanical forces play many important roles in cell biology and animal and plant development.  In contrast to inanimate matter, forces in living matter are generated by active and highly regulated processes within and between cells.  The ability to directly measure forces and mechanical stress on the cellular scale within living tissues is critically important for understanding many morphogenetic processes but is a serious experimental challenge. The present work proposes an alternative approach based on the analysis of images that provide a visualization of cell boundaries in two dimensional epithelial tissues. The method 
uses the assumption of force balance within the epithelial layer to infer, on the basis of image-derived geometric data, the mechanical state of each cell.  The proposed Mechanical Inverse method is illustrated  on the analysis of two examples: the initial step of the gastrulation process in {\it Drosophila} embryo, and the process of neurogenesis in the developing avian cochlea.

\section*{Introduction}
Genetics and biochemistry are central to all aspects of biological function. Physics is often less recognized but still important at many levels, everywhere from intramolecular to organismal scales. In particular, many important aspects of cell behavior depend directly and indirectly on its mechanical state defined by its interaction with neighboring cells and adhesion to the extracellular matrix \cite{lecuitreview,Huang1999,Foty2005}. Cytoskeletal mechanics and cell-cell adhesion determine geometric properties of cells \cite{Evans1989,lecuitreview,kiehart,Kafer2007}, as well as the dynamics of biological tissues \cite{kiehart,pulsedcontract,vertexcellpacking,feedbackregulation,Hayashi2004,Bertet2004,Bao2005,Koppen2006}. In plants, cells do not move, but the rigidity of cellulose membranes makes mechanical stress an obvious factor for cell division and proliferation \cite{Mirabet2011,Hamant2008}.
It is known that animal cell proliferation also depends on substrate adhesion and the degree of cell confinement \cite{Folkman1978,Huang1999,Huang2000,Wozniak2009,Puliafito2011}. It has also been demonstrated that (stem) cell differentiation is affected by substrate rigidity \cite{Engler2006}. More speculatively, mechanical feedback interactions have been conjectured a role in coordination of growth during development \cite{mechregulation,feedbackregulation,Affolter2007,lecuitreview}. Mechanical transformation of epithelial tissue is of course itself central to many morphogenetic processes: gastrulation \cite{pulsedcontract} and convergent extension \cite{lecuitreview}, to name a few.  Understanding how mechanical state changes in cells orchestrate morphological reorganization of tissues is an open problem and a subject of much current work \cite{pulsedcontract,lecuitreview,vertexcellpacking}

Our present understanding of the role of mechanics as one of the regulatory inputs into the cell is strongly impaired by the difficulty of quantitative characterization of the mechanical state (i.e. stress and deformation) of the cell. Among the available techniques are laser tweezers  \cite{Choquet1997} and "traction force microscopy" \cite{Dembo1999,Wozniak2009} performed on cultured cells. UV laser ablation allows to mechanically perturb tissues \cite{vertexcellpacking,kiehartmeasure,rauzilecuit} on cellular scale with the time-lapse imaging of subsequent relaxation providing information on the mechanical state of the tissue. The ablation approach is widely used on live preps, for example, in the study of Drosophila embryonic development.  Yet, this technique is definitely not a "non-destructive" one. 

On the other hand one of the major recent technical advances in developmental biology is the improvement of live fluorescent imaging.  These provide high quality time lapse movies of developmental processes, including interesting morphological transformations such as gastrulation and convergent extension \cite{pulsedcontract,integrationcontract,rauzilecuit}. The purpose of the present investigation is to explore what  insight into the mechanical state of cells may be gleaned from a quantitative examination of high quality images of the type shown in Fig. 1A.
Our goal is to use image analysis as a non-destructive approach to obtaining quantitative measures of stress in these systems. Similar strategy has been pursued by the recently proposed "Video Force Microscopy" (VFM) approach by Brodland et al \cite{Brodland2010}.
Our approach will differ from VFM in its assumptions about mechanical state of tissue, in the parameterization of forces and in the way imaging data is utilized.

Below we shall define a general model parameterizing the mechanical state of cells in two dimensional epithelial tissue and provide a computational method for inferring these parameters from the observed geometry of the cell array. We shall study the sensitivity of the proposed Mechanical Inverse (MI) method to errors in cell geometry and identify conditions under which robust inference is possible. We then illustrate the proposed MI method by applying it to the analysis of two different biological processes: cochlear neurogenesis \cite{goodyear} and ventral furrow formation\cite{integrationcontract}.
\section*{Materials and Methods}
\subsection*{Model of epithelial tissue mechanics.}
Our approach is based on the assumption that epithelial monolayers are in an instantaneous mechanical equilibrium, characterized by a static balance of the forces acting at intercellular junctions. The second important assumption is that  epithelial mechanics is dominated by the actomyosin cortices and inter-cellular Adherens Junctions \cite{lecuitreview} both localized at cell boundaries which form a visible two-dimensional web, as shown in Fig. 1a. Thus we assume that mechanical state of the cell can be described by effective tension at the interface and the hydrostatic pressure in cell interior. Yet, because cells can independently regulate their mechanical state, e.g. by modulating myosin activity or cell-cell adhesion, we allow for the possibility of each intercellular interface to have a different effective tension, $T_{ab}$, and for each cell to have a different internal pressure $P_a$ (where $a$ labels cells and $ab$ labels the interface between cells $a$ and $b$), as shown in Fig. 1D. Mechanical equilibrium then corresponds to the condition that the forces acting on each ``vertex'' $\vec{r_i}$ (defined as a junction of three cells and therefore of three interfaces) add up to zero.

Let $\vec{r}_i$ and $\vec{r}_j$ be the vertices belonging to the interface $ab$ and let $\vec{r}_{ij} \equiv \vec{r}_j -\vec{r}_i$ be the vector from vertex $i$ to $j$. The force exerted by this interface on vertex $i$ is
\begin{equation}
F_{ij}^{\alpha }
= T_{ab} {r_{ij}^{\alpha} \over |r_{ij}|}
+ \frac{1}{2} (P_a - P_b) r_{ij}^\beta  \epsilon_{\beta\alpha},  \label{vforce}
\end{equation}
where $\alpha$ labels vector components in the $xy$ plane and  $\epsilon_{\beta\alpha}$ is the anti-symmetric tensor ($\epsilon_{xy}=-\epsilon_{yx}=1$ and $\epsilon_{xx}=\epsilon_{yy}=0$). As shown in Fig.2, this expression accurately represents the Young-Laplace balance between interfacial tension and  the pressure differential across the interface $P_a - P_b =\kappa_{ab} T_{ab} $, as long as the interfacial curvature $\kappa_{ab}$ is small. This fact enables us to formulate all mechanical balance conditions in terms of a polygonal approximation of the cell array, thus allowing us to reduce the problem to a generalized "vertex model" \cite{vertexcellpacking,feedbackregulation}.


Remarkably, the forces given by (\ref{vforce}) correspond to the mechanical energy in the form of the following simple Hamiltonian
\begin{equation}
 H(\{ \vec{r}_i \}) = \displaystyle\sum_{a}  H_a [A_a, \{ \ell_{a b} \}] \label{Hamil} \end{equation}
where $A_a$ is the area of cell $a$, $\ell_{ab}=|r_{ij}|$ is the length of the interface between cells $a$ and $b$ and  $\{ \ell_{a b} \}$ denotes  the set of interfaces belonging to cell $a$. Both $A_a$  and $\ell_{ab}$'s are defined in the polygonal approximation.   This Hamiltonian is a generalization of the vertex models often used to describe epithelial sheet mechanics  \cite{vertexcellpacking,feedbackregulation,rauzilecuit}.  Pressure and tension are defined by considering the differential form of $H$:
 \begin{equation} \begin{array}{lll}
dH &=& \displaystyle\sum_{ \langle a b \rangle }  \frac{\partial H}{\partial \ell_{ab}} d\ell_{ab} + \sum_{a} \frac{\partial H}{\partial A_a} dA_a  \\
\\
&=& \displaystyle\sum_{ \langle a b \rangle }  T_{ab} d\ell_{ab} + \sum_{a}P_a dA_a \end{array} 
\label{diffHamil} \end{equation}
where we have define $T_{ab} \equiv \frac{\partial H}{\partial \ell_{ab}}$ and $P_a \equiv \frac{\partial H}{\partial A_a}$.  The $\langle a b \rangle$ sum runs over all edges, i.e. pairs of neighboring cells $a$, $b$. This tangent representation of mechanical energy expresses interfacial tension $T$ and intracellular pressure $P$ as conjugate variables to edge lengths and cell areas respectively.  


Mechanical equilibrium means that $H$ is minimized with the respect to vertex positions
\begin{equation}
 F_i^{\alpha} = \sum_{\{ j \}_i } F_{ij}^{\alpha} = - {\partial H(\{ \vec{r}_i \}) \over \partial r_i^{\alpha}} =0 \label{FH} \end{equation}
which defines the static force balance constraints. More generally, the dynamics of passive relaxation towards this equilibrium would be  described by $  \mu { d \over dt} \vec{r}_i = - {\partial H(\{ \vec{r}_i \}) \over \partial \vec{r}_i}$, where $\mu$ is the "effective friction" constant. Our analysis will be based on the assumption that the cell layer is close to mechanical equilibrium in the sense of 
$|\vec{F}_i| << \sum_{\{ j \}_i } |\vec{F}_{ij}|$ meaning that the most of internal forces acting within the tissue are balanced.  If there exists an unbalanced force that drives the physical motion, it is small in comparison to the forces that are balanced.

\subsection*{The mechanical inverse problem.} We can now inquire to what extent the knowledge that a given cell array geometry is in a mechanical equilibrium constrains the parameters $P_a$, $T_{ab}$ describing the mechanical state of cells. We proceed by
a simple count of mechanical constraints and of the free parameters for two cases i) a closed cell array, shown in Fig. \ref{domains_fig}A and ii) an open cell array, shown in Fig. \ref{domains_fig}B.


Let us begin with the closed cell array and let $v,e,c$ to be  respectively the number of vertices, edges, and cells.  For $v$ vertices two dimensions, we have exactly $n_c=2v-3$ mechanical constraints, where the extra three degrees of freedom are associated with global translation and rotation symmetries (alternatively, three constraints are redundant because the total force and total torque in the closed system are equal to zero).  On the other hand, the number of unknown tension parameters is $e$, and the number of unknown pressures is $c$, so that the total number of parameters is $n_p=e+c$.  
Our closed system, if we count exterior as an additional "cell", is topologically equivalent to a sphere so that Euler's theorem reads \Bnum
v - e + (c+1) = 2. \Enum
Combining this relation with the condition that vertices are points where three edges meet and each edge impinges on two vertices, that is $3v = 2e$, we obtain the result
\Bnum \label{counting}
e+c=2v+1. \Enum 
This implies $n_p=n_c+4$, which means that our unknown parameters can be determined up to four free constants. One of the latter is the arbitrary overall scale of $T_{ab}$ and $P_a$ which cannot be constrained by the force balance conditions. (Note also since $P_a$ is only defined up to an additive constant, one can set the pressure in the exterior  of the domain to zero.)  Yet the good news is that the number of free constants is finite, while the number of nontrivial constraints scales with the number of cells!

Repeating the counting procedure for the open system, one finds that $e+c=2v+b+1$, where $b$ is the number of cells at the boundary of the domain.  It follows that $n_p=n_c+b+1$. Thus mechanical parameters are determined up to $b+1$ free constants: we can still choose the overall scale while the additional $b$ degrees of freedom may be regarded as the boundary conditions such as $P_a$'s of the cells at the edge of the domain. Again, for a large array, because $b \sim \sqrt{c}$ while $n_p \sim c$, the number of parameters and constraints is much larger than the number of free constants. 

To actually determine the $T_{ab}$, $P_a$ parameters we use the fact that they appear only linearly in the force balance equations (\ref{FH}) leading to a linear system for 
 \begin{equation} 
 \psi^T = \left(  T_1,  \dots, T_e, P_1, \dots, P_c  \right)
 \label{psi} 
 \end{equation}
 in the form
 \begin{equation} 
M\psi = C 
 \label{configeqn} 
 \end{equation} 
 with $M$ being an $n_p \times (n_c+1)$ matrix the 1st $n_c$ rows of which impose force balance conditions and the additional row imposing the scale, by constraining the average tension to be equal to one. Correspondingly the top $n_c$ entries of the column vector $C$ are zero, while the bottom row $C_{n_c+1}=1$.
 
The rectangular system (\ref{configeqn}) is solved by via pseudo-inverse \cite{pseudoinverse} with the general  
solution  of the form 
\begin{eqnarray}
\psi = \Psi + \displaystyle \sum_{\nu=1}^{n_z} A_\nu \phi^\nu 
\label{fullsoln1} \end{eqnarray}
with
\begin{eqnarray}
&\Psi =\tilde{M}^{-1}C,\\
&M \phi_\nu = 0 
 \label{fullsoln2} \end{eqnarray}
with $\tilde{M}^{-1}$ being the pseudo-inverse of the rectangular matrix $M$ and the amplitudes $A_\nu$ of the $n_z=n_p-n_c-1$ "zero modes" $\phi_{\nu}$ are the free parameters.

Fixing the remaining $n_z$ degrees of freedom requires introducing additional constraints: e.g. one may have reasons to seek a solution which minimizes variation of $P_a$'s or $T_{ab}$'s. In choosing such additional assumptions one may want to use all the information that one has for specific applications, as we shall do below. However, before proceeding to the applications we must consider the issue of error sensitivity. 
\subsection*{Sensitivity of the inverse.}  Our  approach to mechanical parameter inference is based on the observed geometry of the cell array. How sensitive are the results to the inaccuracy of vertex positions $\{ \vec{r}_i \}$? Such an inaccuracy will inevitably arise in the process of imaging and image segmentation and even more importantly from the fact that cells fluctuate and our assumption that any particular configuration is in equilibrium, is at best approximate. To quantify the stability of the inverse we consider the effect of an arbitrary small perturbation in vertex positions, $\{ \delta \vec{r}_i \}$. Because the inhomogeneous term $C$ in (\ref{configeqn}) is independent of cell geometry, the variation of parameters $\delta \psi$ in the response to positional error  is given by
\Bnum
&M\delta\psi +\left[  \frac{\partial M}{\partial \vec{r}} \psi \right] \delta \vec{r} =0 \\
&\delta\psi =L \delta\vec{r} = \left[\tilde{M}^{-1} \frac{\partial M}{\partial \vec{r}} \psi \right] \delta\vec{r}. \label{response_m}\Enum


Ideally the error response matrix $L$ has small eigenvalues providing a relatively robust inverse.  On the other hand, large eigenvalues of $L$ would indicate high error sensitivity. These sensitive modes appear via the pseudoinverse matrix $\tilde{M}^{-1}$. A histogram of singular values of the matrix $\tilde{M}^{-1}$ is shown in blue in Fig. \ref{errordensity} (for a closed system with $\approx 800$ cells). One notes that a substantial fraction of modes have eigenvalues larger than one.  As a result, small errors in positions can result in large error in inferred parameters. 

The simplest way to solve the sensitivity problem is to reduce the number of parameters. For example, as we shall argue below, in some contexts it may be reasonable to neglect variation in cell pressure and set $P_a=P_0$ which eliminates $c$ parameters, reducing $n_p$ from $4c$ to $3c$. In that case the mechanical constraint system given by (\ref{configeqn}) becomes overdetermined and can be solved only in the sense of least square minimization: i.e. minimization of
 \begin{equation}
Tr[(M'\psi - C )^T(M'\psi - C)].
 \label{leastsq}
 \end{equation}
The solution of the minimization problem is still given by the pseudo-inverse of the reduced rectangular matrix $M'$, the reduction being accomplished by eliminating constrained unknown parameters from in $\psi$. Fig. \ref{errordensity} shows (in red) the distribution of singular values governing the sensitivity of the reduced or {\it partial} inverse problem. We note a substantial reduction in sensitivity.
 
The partial inverse approach can tested {\it in silico}.  To that end we consider a closed array of cells that appears in Fig. \ref{errordensity}A and define cell geometry by minimizing elastic energy given by 
\begin{equation}
H_V (\{ \vec{r} \} )= \sum_{\bra ab \ket } k_{ab} (\ell_{ab} - 1)^2.
\end{equation}
with uniformly distributed $k_{ab} \in [0.7,1.3]$.  The absence of area terms imposes constant pressure.  (The cell array is relaxed under toroidal boundary conditions to prevent a collapse into the zero tension ground state.)
The vertex model parameters are computed via equation (\ref{Hamil}).  These quantities are then compared to values obtained by applying the partial inverse algorithm to the vertex ``data'' $\{ \vec{r}_i \}$ corrupted by random noise $\{\delta \vec{r}_i\}$ with an r.m.s. variation of 5\% of the average length of cell edge (see Fig. (\ref{part_inv_5pct})).  The correlation coefficient between inferred and computed parameters is 0.852, which confirms the ability of our method to extract information from noisy data. 


We note that the "soft modes" which give rise to the sensitivity of the full inverse problem are quite interesting.  The formulation of the minimally constrained problem is analogous to the isostatic systems studied in jamming transitions of amorphous solids \cite{wyart}.  These isostatic systems live on the boundary of Maxwell's criterion for rigidity, and much like amorphous solids, they must satisfy both the local and global rigidity conditions.  In our mechanical inverse formulation, "rigidity" corresponds to a fully constrained set of mechanical ($T_{ab}$ and $P_a$) parameters. Amusingly, local soft modes for the MI problem correspond to special local geometries: specifically, polygons that can be inscribed into circles (i.e. generalization of regular polygons) - a category which includes triangles of any shape. These interesting mathematical aspects of the problem will be discussed in a separate publication.
\section*{Results}
\subsection*{Mechanical differentiation of cells in the developing avian cochlea.}
During cochlear development, which takes place during the 1st two weeks of chick embryonic development, cells in the initially homogeneous two dimensional epithelial layer differentiate into pro-neural (hair-cell) and support cell fates \cite{goodyear}. The process is driven by the Delta/Notch-mediated cell-contact signaling \cite{bray2006} which causes Òlateral inhibitionÓ: cells which express Delta ligand on their surface prevent their immediate neighbors from doing the same. Expression of Delta  is an early marker of the pro-neural fate of cells. Fig. 1A presents an image of the cochlea epithelium, obtained by Goodyear and Richardson \cite{goodyear} at the stage of development shortly after the onset of differentiation. The two cell types already have a discernibly different morphology: pro-neural cells are somewhat smaller and have curved edges. This dimorphism is supported by direct labeling of specific pro-neural markers, shown in Fig. 1B and demonstrated in \cite{goodyear}. 

Our goal is to infer, based on the analysis of the image in Fig. 1A, the variation in the mechanical parameters between cells. The visible positive curvature associated with pro-neural cells suggests that they are under higher internal pressure. Can the Mechanical Inverse method determine pressure differentials between cells? Because our approach requires only positions of cellular vertices, it does not use the information provided by the  interfacial curvatures which are readily measurable on the image. This additional information will be used as an {\it a posteriori} validation of the inferred results. 

\begin{figure}[!ht]\begin{center}
\includegraphics[width=0.4\textwidth]{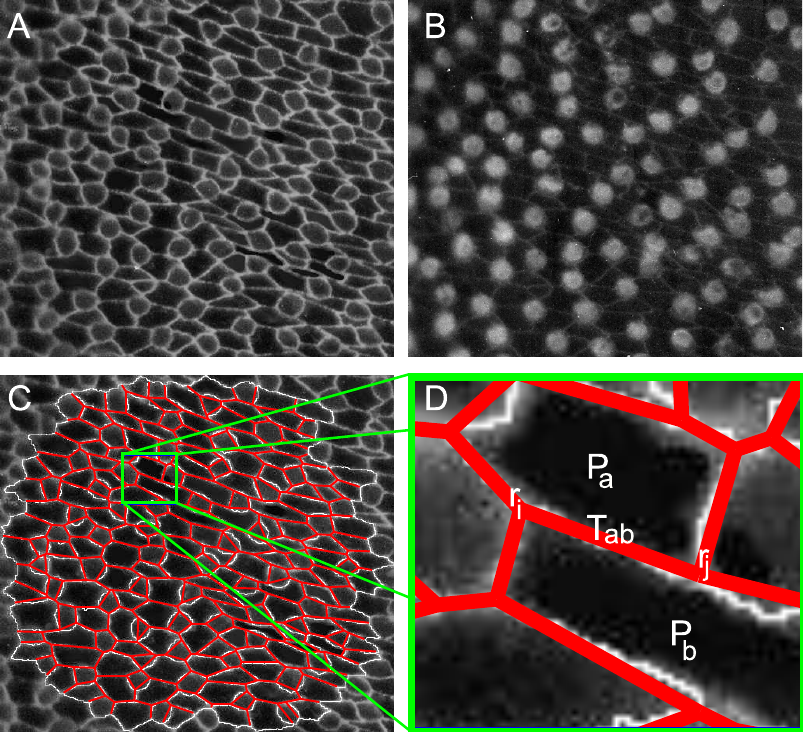}
\end{center}
\caption{Avian cochlear epithelium {\bf(A)} at the E9 stage of development just following differentiation of cells into hair-cell precursors and surrounding support cells (from Goodyear and Richardson \cite{goodyear}). Panel {\bf(B)} shows the same tissue with stained (white) pro-neural cells.   Panel {\bf(C)} shows a computer generated segmentation of the raw image in {\bf(A)} as a polygonal tiling which approximates cell geometry.  The zoomed-in image {\bf(D)} defines our parametrization of cell geometry in terms of vertex coordinates $r_i$ and of the mechanical state of the cell in terms of interfacial tensions $T_{ab}$ and hydrostatic pressures $P_a$.}
\label{intro_vertex_fig}\end{figure}

To reduce the number of parameters we assume that interfacial tensions can be expressed as $T_{ab} =T_{a} +T_b$ in terms of Òcortical tensionsÓ $T_a$, $T_b$ of adjacent cells. This reduces the number of parameters by $e-c=2c$ which is sufficient to render a robust partial inverse (in the sense of least squares), yielding $T_a$ and $P_a$ for every cell. Fig. \ref{27_01_pa_tl} shows the distribution of inferred intracellular pressures and cortical tensions, for the two cell types. We see that pro-neural cells have on average higher tension and pressure. While pressure shows some correlation with cell area, there is no correlation between interfacial tension and its length. There, however is no reason to expect any specific correlation between these quantities. On the other hand, Laplace' Law predicts $ P_{a}-P_{b} = \kappa_{ab} T_{ab} $ which we are in a position to check directly, thanks to the fact that interfacial curvatures $\kappa_{ab}$ are directly measurable on the images such  Fig. \ref{intro_vertex_fig}A.
Fig. \ref{dP_corr} presents the "empirical" Laplace' Law obtained on the basis of the inferred $ P_{a}-P_{b}$ and $T_{ab} $. Because the Mechanical Inverse algorithm did not in any way use the interfacial curvature information, the fact that inferred parameters approximately obey the Laplace' Law provides a validation of  the inverse method.




\subsection*{Mechanical anisotropy at the onset of the ventral furrow formation in Drosophila.}  Ventral furrow formation in Drosophila is the first step of the gastrulation process and begins with the contraction of apical surfaces of cells along the ventral midline of the ellipsoidal monolayer of cells that comprise the embryo at that early stage of development \cite{Wolpert2002,Gilbert2003, integrationcontract}. Fig. 10A presents the ventral view of a Drosophila embryo at the beginning of this mechanical transformation. The high quality of these images makes it possible to attempt the Mechanical Inverse analysis.  Since the process begins even before cellularization is completed it is reasonable to assume that cells have the same internal pressure $P_a =P_0$, allowing us to reduce the number of parameters enough to achieve a robust partial inverse and infer $T_{ab}$ for every cell boundary. We find a rather broad distribution of tensions (with the coefficient of variation $\approx 0.3$).  

Interestingly, comparing images separated by merely two minutes (Fig. \ref{anisotropy_comphist}) we found that the inferred $T_{ab}$ at the later time-slice exhibited statistically significant anisotropy with estimated tensions of cell interfaces along the AP axis being on average about 15\% higher than those along the DV axis. The inferred increase in AP tension (relative to DV) is consistent with the laser ablation measurements made in the Wieschaus lab \cite{pulsedcontract,integrationcontract}. Yet, mechanical inverse inference gives information not only on the global, tissue-wide level, but also on the scale of a single cell and interface. The analysis also clearly demonstrates the ability to make specific predictions (for interfacial tensions) that can be directly tested by combining high quality live imaging with UV pulsed laser ablation.



\subsection*{Intercellular traction forces.}  The variation of tension from one interface to another implies the existence of traction forces acting between cells. This traction, or shear stress, must be entirely borne by the cadherins which bridge cellular membranes and connect actomyosin cortices of apposing cells  \cite{lecuitreview}. In Fig. \ref{coarse_stress_mod} we zoom in on an interface decomposing interfacial tension into the cortical tensions on the opposite sides of the interface $T_{ab}=T_a+T_b=T_{a'}+T_{b'}$, allowing for the possibility that the latter are not constant along the interface and vary as a function of position along the edge. (Here,  $T_a$ and $T_{a'}$ refer to the two ends of the edge as shown in Fig. \ref{coarse_stress_mod}.) This transfer of tension from the cortical bundle in one cell to the other is possible because of cadherin mediated traction forces acting between cells. The total shear stress on the interface is $\tau_{ab}= (T_{a'}-T_a)/{L_{ab}}=(T_b-T_{b'})/{L_{ab}}$. In the Supplementary Information we show that because cortical tensions are constrained by the continuity conditions at cell "corners" they can be readily expressed in terms of interfacial tensions leading to the following simple expression for the traction force acting between cells $a$ and $b$. 
 \begin{equation}
 \tau_{ab}= \frac{1}{L_{ab}}(T_{ac}-T_{ad}+T_{bd}-T_{bc})
 \label{traction}
 \end{equation}
 Fig. \ref{E1_imgoverlay}B shows inferred tractions calculated for the ventral furrow data. We observe a significant variability in tractions at different interfaces. Because traction forces stretch trans-cellular cadherin dimers, they may be physiologically important. Since at present there is no way of measuring them directly  the possibility of indirect inference is particularly interesting. 

\section*{Discussion}
We have demonstrated that the readily visualized two dimensional network of cellular interfaces in an epithelial tissue holds, potentially, a wealth of information on the relative strength of mechanical stresses acting in the tissue. The main precondition is that the tissue is close to the mechanical equilibrium in which internal cytoskeletal forces are balanced by intercellular interactions. Any imbalance of forces corresponding to directed or fluctuating motion must be small by comparison with the balanced static component. Force balance is achieved by the suitable adjustment of cell geometries (parameterized by the positions of vertices). Conversely we envision changes in tissue geometry to be driven adiabatically - i.e. without disruption  of the mechanical equilibrium - by changes in cytoskeletal forces within cells. This picture is at once similar and dissimilar to the case of soap froths. The geometry of a soap froth \cite{glaziergraner,stavans,weaire} is also defined by the instantaneous force balance and changes adiabatically (when gas diffuses out of cells with higher internal pressure). Yet  epithelial cells, in contrast to soap bubbles, can control interfacial tension by regulating myosin activity within actomyosin cortices and therefore can generate variation in tension on sub-cellular scale, even between different interfaces of the same cell. 

Our Mechanical Inverse method is fundamentally different from the Video Force Microscopy \cite{Brodland2010}. In contrast to our assumption that cytoskeletal forces are in an approximate instantaneous balance, VFM is based on the assumption that bulk forces acting within the tissue are balanced by viscosity: inverse is therefore based on the observed velocity of tissue motion. VFM employs finite element methods to define forces on a computational grid rather than underlying cells. The two methods are complementary in the sense that VFM provides information about the distribution of unbalanced bulk force which drives motion on the scale of the embryo, while our Mechanical Inverse is focused on the internal balance of forces in relation to cell geometry and its local changes.

The proposed Mechanical Inverse method, converts clearly stated assumptions about the nature of cellular stresses into readily falsifiable predictions. 
 Using the example of avian cochlea, we were able to demonstrate that mechanical parameters inferred via the Mechanical Inverse satisfy non-trivial cross-checks provided by independent additional information (interfacial curvature measurements) read off the tissue images. Thus our approach is capable, in realistic applications, to infer mechanical parameters and to uncover interesting aspects of the internal state of the cell. By combining high quality live imaging with UV pulsed laser ablation, one will be able to put predictions for local interfacial tensions obtained via the Mechanical Inverse, to a rigorous experimental test. We note however, that the predictions do not have to be very accurate, to be useful. Even if inferred tensions each carry only one bit of information - i.e. identify interfaces with high or low tension - correlating tension with the observed level of myosin, cadherin and/or other proteins involved in regulation of cell mechanics could be extremely informative. (Since a large number of cells can be imaged and analyzed, the method is effectively "high throughput"!) In addition our approach
 allows  to infer quantities such as inter-cellular traction forces (or shear stress), which may well be important for the stability of Adherens Junctions but cannot be directly measured by any presently available means. Hence we expect that further development, validation and application of the Mechanical Inverse method will leads to new insights into the molecular biology of epithelial cells and tissues.

\section*{Acknowledgments}
The authors acknowledge stimulating discussions with M. Kaschube, T. Lecuit M. Mani, D. Sprinzak and E. Wieschaus and thank
R. Goodyear and Wieschaus' Lab for providing the images used in our analysis. This work was supported by NSF PHY-0844989.

\bibliography{biblio}

\section*{Figure Legends}

\begin{figure}[!ht]\begin{center}
\includegraphics[width=0.4\textwidth]{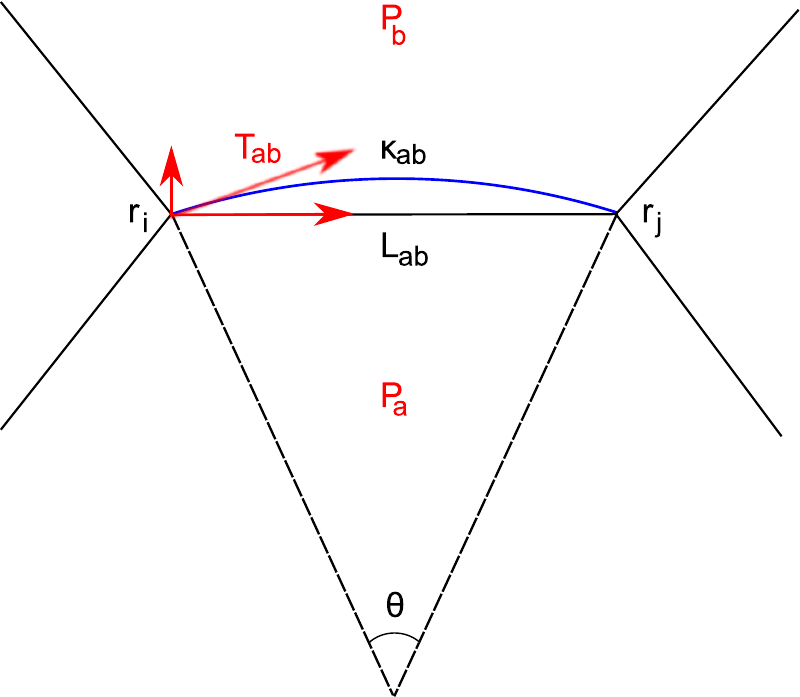}
\end{center}
\caption{Schematic representation of an edge  between two cells in the tissue, comparing a curved interface (blue) with its approximation by a chord defining the edge in the polygonal representation of cells.  Mechanical stress parameters are in red, and geometric quantities are labeled in black.  Provided that the curvature of the interface $\kappa_{ab}$ (and hence the angle $\theta$) is small, the Young-Laplace equation $P_a - P_b = \kappa_{ab} T_{ab}$ defines the force on the vertex $i$ between cells $a$ and $b$ which obeys Eqn. (\ref{vforce}).}
\label{younglaplace_relation}\end{figure}

\begin{figure}[!ht]\begin{center}
\includegraphics[width=0.4\textwidth]{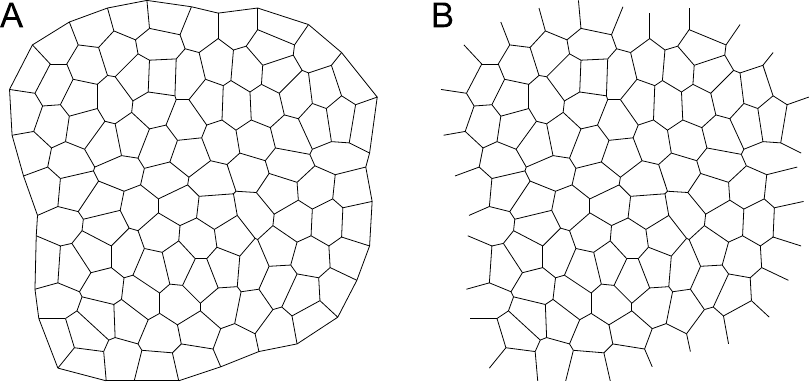}
\end{center}
\caption{Examples of computer generated {\it closed} {\bf(A)} and {\it open} {\bf(B)} cell arrays. Closed arrays provide an idealized context for defining the mechanical inverse problem, while the analysis of experimental data requires dealing with open arrays, corresponding to convex patches of cells defined by or within the field of view.}
\label{domains_fig}\end{figure}

\begin{figure}[!ht]\begin{center}
\includegraphics[width=0.4\textwidth]{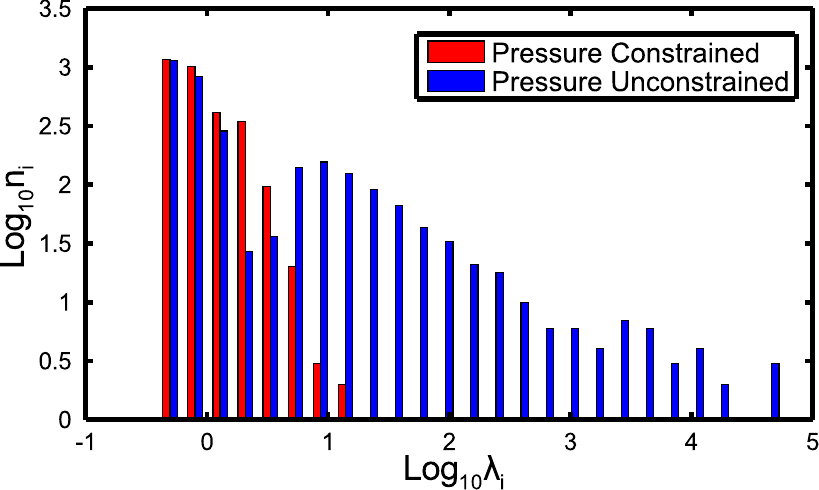}
\end{center}
\caption{The distribution of singular values (which correspond to the square root of non-zero eigenvalues of $[M^TM]^{-1}$) for the $\tilde{M}^{-1}$ matrix before (blue) and after (red) parameter reduction.  Note that prior to parameter reduction there are a substantial fraction of eigenvalues $>10$  which means that small errors in vertex positions are significantly amplified in solving the inverse problem. Large eigenvalues are effectively suppressed after parameter reduction.} \label{errordensity}\end{figure}

\begin{figure}[!ht]\begin{center}
\includegraphics[width=0.4\textwidth]{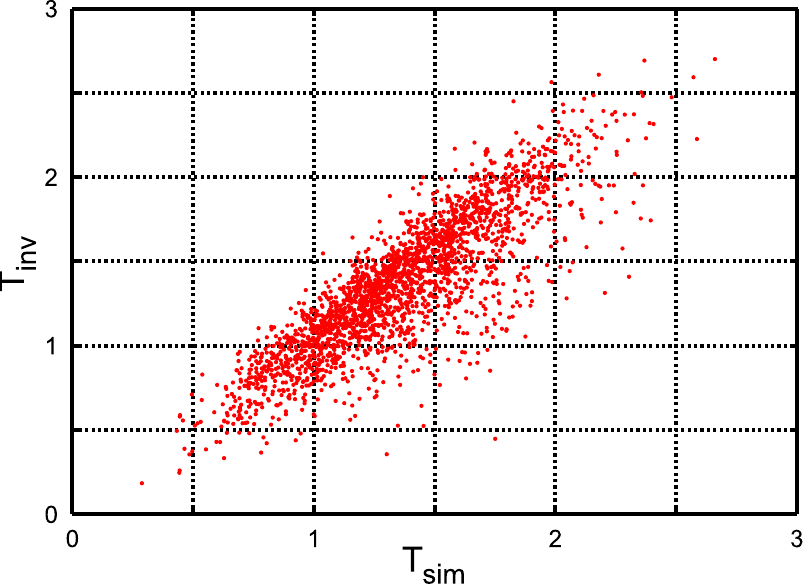}
\end{center}
\caption{Scatter plot comparing actual  $T_{ab}$ describing the {\it in silico} cell array
 to the values inferred by the partial inverse algorithm applied to the vertex data corrupted by 5\% random noise. The plot exhibits a correlation coefficient of 0.85 between the estimated and actual tensions.}  \label{part_inv_5pct}\end{figure}
 
\begin{figure}[!ht]\begin{center}
\includegraphics[width=0.4\textwidth]{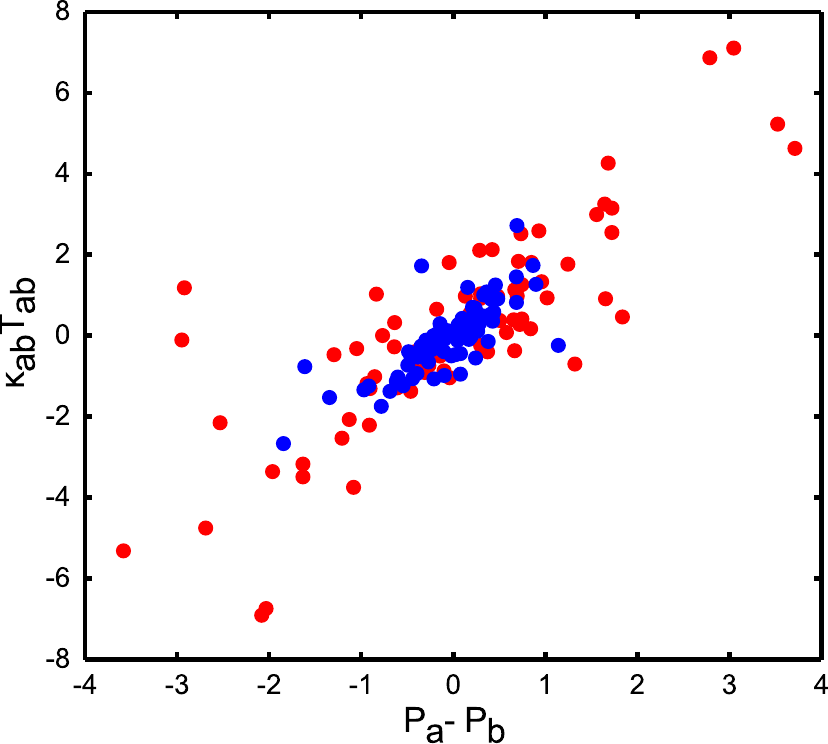}
\end{center}
\caption{Scatter plot comparing inferred pressure differential across an interface, $ P_{a}-P_{b} $, with the product of inferred tension $T_{ab}$ and the measured curvature $\kappa_{ab}$ of the same interface. Different colors distinguish results obtained from different images.
The scatter plot exhibit a clear correlation between the two quantities, as expected from the Laplace' law $ P_{a}-P_{b} = \kappa_{ab} T_{ab} $.}
\label{dP_corr}\end{figure}

\begin{figure}[!ht]\begin{center}
\includegraphics[width=0.4\textwidth]{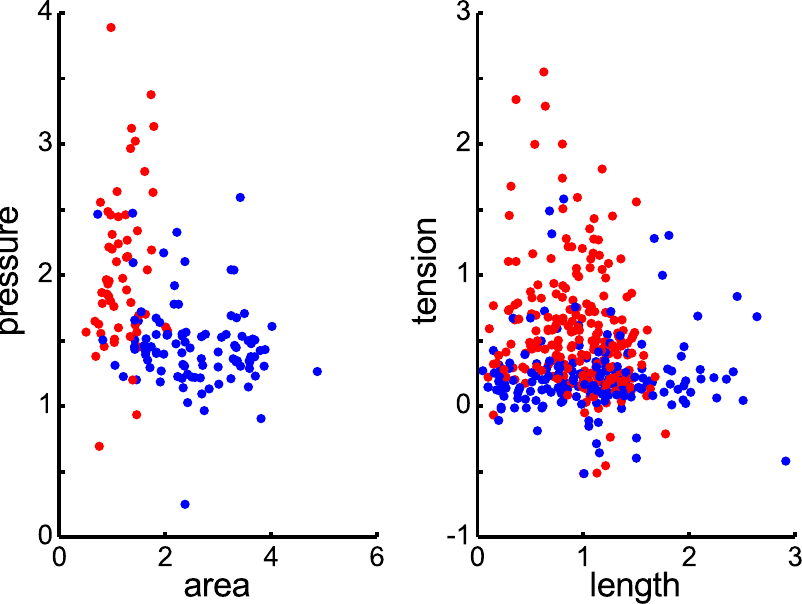}
\end{center}
\caption{Inferred tensions and pressures for the cochlear epithelium image shown in Fig. 1A. Hair-cell precursors and support cells correspond to red and blue dots respectively. Inferred pressure is plotted versus observed cell area and inferred tension versus edge length. Note systematically higher inferred pressure and tension in the hair-cells.}
\label{27_01_pa_tl}
\end{figure}

\begin{figure}[!ht]\begin{center}
\includegraphics[width=0.4\textwidth]{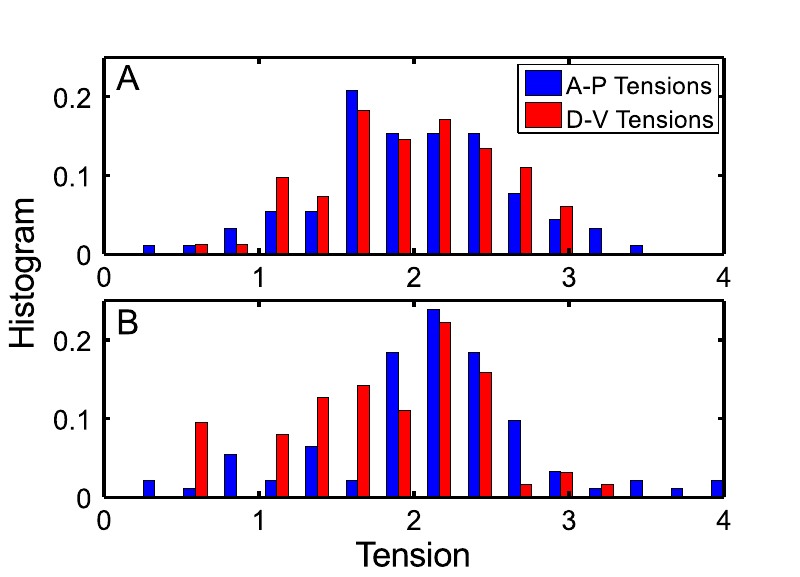}
\end{center}
\caption{Histograms of inferred tension at the start of the ventral furrow formation. Red (blue) corresponds to cell edges at an angle above (below) $\theta_c = \pi/4$ relative to the AP axis.  Panels {\bf (A)} and {\bf (B)} correspond to respectively the 1st and the 3rd minutes of  the furrow formation process.  }
\label{anisotropy_comphist}\end{figure}
 
\begin{figure}[!ht]\begin{center}
\includegraphics[width=0.4\textwidth]{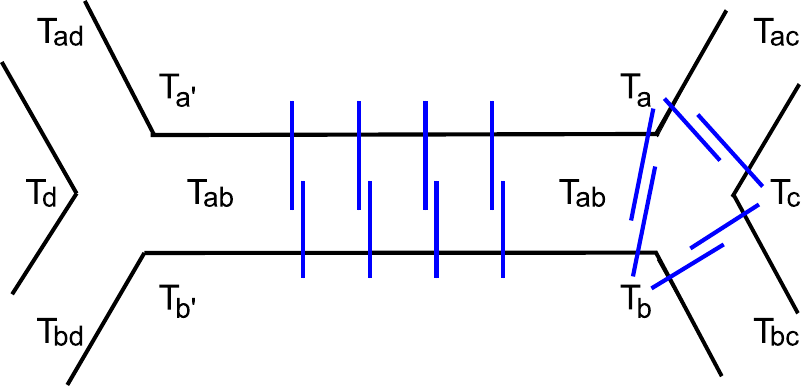}
\end{center}
\caption{Schematic decomposition of the effective interfacial tension  into cortical tensions acting within cells: $T_{ab} = T_a + T_b$, $T_{bc} = T_b+T_c$, etc.  Because cytoskeletal cortexes of cells are crosslinked by cadherins via Adherence Junctions, indicated in blue, cortical stress can be transferred from one cell to another so that $T_a \neq T_{a'}$. The corresponding traction force (or shear stress) is given by Eqn. (\ref{traction}).} \label{coarse_stress_mod}\end{figure}
 
\begin{figure}[!ht]\begin{center}
\includegraphics[width=0.4\textwidth]{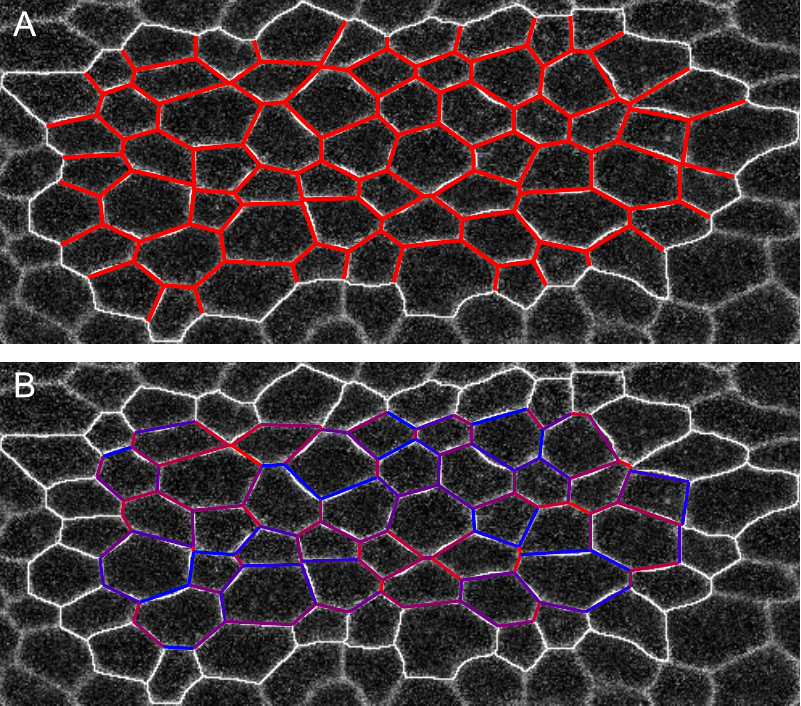}
\end{center}
\caption{Images of the ventral side of a Drosophila embryo 4min prior to ventral furrow invagination \cite{integrationcontract}. Panel {\bf (A)} shows the polygonal tiling array defined by image segmentation.  Panel {\bf (B)} shows inferred tractions obtained from the partial inverse and Eq. (\ref{traction}). Color indicates the magnitude of inferred traction with red (blue) being the relatively high (low) traction.  The coefficient of variation of inferred traction is $\approx 0.2$.}
\label{E1_imgoverlay}\end{figure}
 

\end{document}